\documentclass[12pt,preprint]{aastex}
\usepackage{epsfig,psfig}

\newcommand{\be}{\begin{equation}}
\newcommand{\ee}{\end{equation}}
\newcommand{\bea}{\begin{eqnarray}}
\newcommand{\eea}{\end{eqnarray}}

\def\gsim{\lower.4ex\hbox{$\;\buildrel >\over{\scriptstyle\sim}\;$}}
\def\lessim{\lower.4ex\hbox{$\;\buildrel <\over{\scriptstyle\sim}\;$}}

\slugcomment{to be sumitted to Astrophys. J.}

\shorttitle{Depolarization of solar radio bursts}
\shortauthors{Melrose}

\begin{document}                                                                                   

\title{Depolarization of radio bursts due to reflection off sharp boundaries in the solar corona}

\author{D. B. Melrose}
\affil{School of Physics, University of Sydney, NSW 2006, Australia}

\begin{abstract}
It is argued that depolarization of solar radio bursts requires reflection off boundary layers no thicker than about a wavelength (a few meters at most) between regions with large density ratios. The implied inhomogeneities suggest that the corona is much more highly and sharply structured than can be resolved from observations at other wavelengths. A simplified version of magnetoionic theory is used to derive a depolarization coefficient: the effect of the magnetic field is ignored in treating the dispersion, but taken into account in treating the (circular) polarization. Plots of the depolarization coefficient are used to infer conditions under which effective depolarization occurs, and it is concluded that favorable conditions require total internal reflection. For type~I sources away from the central meridian, effective depolarization requires reflection off an overdense structure with a density ratio $\xi\approx10$. For type~III bursts, a density ratio $\xi\approx2$ suffices, with at least two reflections off walls of ducts at $\approx20^\circ$ to the radial.
\end{abstract}


\section{Introduction}

There is compelling evidence for a depolarizing agent operating on solar radio emission as it propagates through the solar corona. The evidence is from depolarization of fundamental (F) plasma emission, which simple theory predicts should be 100\% in the sense of the o~mode \citep{M80,M85,M91}. This prediction is supported by observations of type~I emission, which is close to 100\% polarized in the sense of the o~mode for sources near the central meridian \citep{E77}. However, type~I sources closer to the limb have an intermediate polarization that decreases as the source approaches the limb \citep{Z75,WZM86}.   A model for the directivity of type~I emission invokes reflection off overdense fibers \citep{BS77}, for sources away from the central meridian, and depolarization is thought to be associated with this reflection \citep{WZM86}. There is also evidence for intermediate polarization in type~IV emission \citep{CZ95}, indicating that this effect is not unique to type~I sources. Depolarization is required for all F~emission in type~II and type~III bursts. Such F~emission can be moderately to relatively highly polarized ($\approx70$\%) in the sense of the o~mode \citep{DSS84,SD85}. However,  it is never 100\% polarized. If depolarization is due to reflection, this implies that  type~II and type~III  F~emission always experiences at least one depolarizing reflection before escaping from the corona. There is strong evidence for ducting of both F and H (harmonic) type~III emission \citep{D79}, and one suggestion is that the depolarization results from reflection off the walls of a duct \citep{H85b}. 

In this paper it is assumed that depolarization of solar radio bursts is due to reflection off field-aligned, overdense structures with sharp boundaries, and the implications of this assumption are explored. The theory of reflection and transmission of magnetoionic waves at a sharp boundary exhibits a rich and varied range of features \citep{H85a,H85b}. This makes it difficult to identify the general features needed in formulating a semiquantitative theory of depolarization. A relatively simple analytic model for the depolarization is needed. Such a model is proposed here: it corresponds to a limit of  magnetoionic theory in which the magnetic field is assumed to be arbitrarily small. This is a valid approximation in cases where the polarization of the natural modes is approximately circular, and when the difference between the refractive indices of the o~and x~modes is small compared with their mean. The most serious limitation of this approximation is near the plasma frequency, where there is a range of frequencies where the o~mode can propagate and the x~mode is either evanescent or is becomes the z~mode. (This limitation is no important provided that the reflecting boundary is well above the source region, which is the case in the models discussed here.) For incident radiation that is 100\% circularly polarized, this model is used to calculate the reflection coefficients for radiation with the same handedness and the opposite handedness. The net degree of circular polarization of the reflected radiation is identified as the depolarization coefficient. The reflection is assumed to be off a sharp boundary over an overdense region, with a density ratio $\xi$, aligned along magnetic field lines, with the incident ray at an angle $\theta$ to the field line.  

The requirement for reflection to lead to depolarization is discussed in section~2: existing arguments that the boundary can be no more than about a wavelength (a few meters) thick \citep{B61,H85a} are supported by an argument based on mode-coupling theory. The simplified form of magnetoionic theory is introduced in section~3 and used to calculate a depolarization coefficient for reflection off a sharp boundary. In section~4 it is argued that effective depolarization requires total internal reflection, and favorable conditions for depolarization of type~I and type~III bursts are identified. The results are summarized and discussed in section~5.

\section{Reflection versus refraction}

The difference between refraction through a large angle in a boundary layer and reflection at a sharp boundary is discussed in this section. When refraction applies, wave propagation in an inhomogeneous medium may be described using mode-coupling theory, with each mode described by a WKB solution. Reflection applies when the density gradient is so steep that mode coupling theory is invalid. 

\subsection{Qualitative discussion of mode coupling}

If any inhomogeneity is sufficiently weak, radiation initially in a single magnetoionic mode remains in that mode. A small leakage into the other mode is described by the theory of mode coupling. A standard treatment of mode coupling can be summarized as follows \citep{B61,M74a,M74b}. Suppose the plasma density (or some other plasma parameter) increases smoothly along a specific direction, identified as the $z$-axis. The equations of the magnetoionic theory are Fourier transformed in time and in $x$ and $y$, and reduced to four coupled linear differential equations in $z$ for $E_x,E_y,B_x,B_y$. In the homogeneous case, on also Fourier tranforming in $z$, the condition for a solution to exist reduces to a quartic equation (the `Booker quartic' equation) for $k_z$. In simple cases, the four solutions may be interpreted as four different modes and labeled as ${\rm o}\!\uparrow$, ${\rm o}\!\downarrow$, ${\rm x}\!\uparrow$ and ${\rm x}\!\downarrow$, where the arrows refer to the direction of propagation relative to the $z$-axis. Including the inhomogeneity leads to coupling between these four modes. Coupling becomes strong near coupling points, where the values of $k_z$ for two modes are equal. Leakage between one mode and the other is due to mode coupling between ${\rm o}\!\uparrow$ and ${\rm x}\!\uparrow$ (or ${\rm o}\!\downarrow$, and ${\rm x}\!\downarrow$). A reflection point for the o~mode or the x~mode refers to a point along a curved ray path where $k_z$ passes though zero and a refracting ray reverses its sense of propagation along the direction of the gradient. At a reflection point, coupling between ${\rm o}\!\uparrow$ and ${\rm o}\!\downarrow$ (or ${\rm x}\!\uparrow$ and ${\rm x}\!\downarrow$) becomes strong. For the theory of mode coupling to be valid, the coupling points must be well separated, requiring that the coupling between say ${\rm o}\!\uparrow$ and ${\rm x}\!\downarrow$ near the reflection point for the o~mode or the x~mode must be weak for the theory to apply. 

In contrast, at a sharp boundary, an incident wave in one mode produces reflected (and, if allowed, transmitted) waves in both modes. Mode coupling theory cannot be valid when reflection leads to two reflected modes, which is the case of interest in explaining depolarization. The transition from reflection-like refraction to true reflection cannot be treated within the framework of mode-coupling theory, and requires a full-wave theory \citep{B61,H85a}.

\subsection{Separation of reflection points}

When the density gradient is sharp, reflection and transmission are treated by applying the electromagnetic boundary conditions at the interface. This requires that the reflection points, in the sense used in mode-coupling theory, be effectively coincident. This qualitative condition can be converted into a semi-quantitative condition for reflection as follows.

Given an analytic model for the boundary layer, geometric optics allows one to determine the location of the reflection points for the o~and x~modes inside the boundary layer.  In the vicinity of a reflection point, the full wave solution may be approximated by an Airy function \citep{B61}, which is an oscillating function on one side and an exponentially decaying function on the other side of the reflection point. At the reflection point for the x~mode, which is on the lower-density side of the reflection point for the o~mode, the x~mode becomes evanescent. For reflection to apply in this case, the skin depth of the x~mode beyond its reflection point must be large in comparison with the distance between the reflection points for the x~mode and the o~mode. Only then can radiation tunnel from one reflection point to the other before decaying significantly. The way this skin depth is estimated is outlined in Appendix~A, cf.\ (\ref{Imq}). Let the density gradient be characterized by a distance $L=1/|{\rm grad}\,n_e|$. The requirement for reflection to apply, when the cyclotron frequency is much smaller than the plasma frequency, $\Omega_e\ll\omega_p$, then becomes
\be
L\lessim(c/\omega)[(1-r^2)(1+r^2\cos^2\phi)]^{1/2},
\label{reflection}
\ee
where $\phi$ is the angle between the incident ray direction and the plane defined by the density gradient and ${\bf B}$. The parameter $r$ may be interpreted as the sine of the angle of incidence times the initial refractive index. It follows that the plasma frequency much change over a distance of order the free-space wavelength in order for reflection rather than refraction to apply \citep{B61,H85a}.

\section{Depolarization due to reflection}

Depolarization due to reflection off a sharp boundary is due to incident radiation in one mode resulting in reflected radiation in both modes. A simplified model is used here to derive a formula for the depolarization on reflection. 

\subsection{Simplified model for reflection coefficients}

The approximate treatment of depolarization due to reflection is based on the following assumptions.
\begin{enumerate}
\item The magnetic field is neglected in the refractive indices for the magnetoionic modes but is taken into account in their polarizations, which are assumed circular, denoted $R,L$.
\item The boundary layer is parallel to the magnetic field lines.
\item The incident wave is in the plane defined by the magnetic field lines and the normal to the boundary layer.
\end{enumerate}
The following parameters need to be specified. The ratio of the wave frequency to the plasma frequency on the low-density side is specified by $X=\omega_p^2/\omega^2$, and the corresponding ratio on the high-density side by $X'=\omega'^2_p/\omega^2=\xi X$. The angle of incidence also needs to be specified, and it is convenient to write this as $\theta_{\rm inc}=\pi/2-\theta$, where $\theta$ is the angle between the incident ray and the magnetic field. Let the incident ray be in a plane at an azimuthal angle $\phi$. In general, there is a transmitted ray, at angles $\theta',\phi'$ and a reflected ray at $\theta'',\phi''$. These satisfy Snell's law in the form
\be(1-X')^{1/2}\cos\theta'=(1-X)^{1/2}\cos\theta, 
\quad
\phi'=\phi, 
\qquad
\theta''=\theta, 
\quad
\phi''=\phi+\pi.
\label{Snell}
\ee
Assumption  3 corresponds to $\phi=0$, with the incident, reflected and transmitted waves all in the plane containing the direction of the gradient and the magnetic field. Then the reflected ray is deflected through an angle $\theta_{\rm def}=2\theta$.

In this model, there are only two different reflection coefficients: $r_{RR}=r_{LL}$ and $r_{RL}=r_{LR}$, where the first (second) subscript indicates the polarization of the incident (reflected) mode. Assuming incident $R$-polarization, the depolarization coefficient is
\be
p_R={r_{RR}-r_{RL}\over r_{RR}+r_{RL}}.
\label{pR1}
\ee
Calculation of the Fresnel coefficients in this model involves modifying the standard textbook derivation by separating the incident (unprimed), transmitted (primed) and reflected (double primed) waves into circularly polarized components, $E_{R,L}$, $E'_{R,L}$, $E''_{R,L}$. For incident $R$ polarization one sets $E_L=0$, and solved for the Fresnel coefficients $E''_{R,L}/E_R$, with $r_{RR}=|E''_{R}/E_R|^2$ $r_{RL}=|E''_{L}/E_R|^2$. The total reflection coefficient is $r_{\rm tot}=r_{RR}+r_{RL}$, and the transmission coefficient is $1-r_{\rm tot}$. 

Three different cases need to be distinguished in writing down the Fresnel coefficients:
\begin{description}
\item[A.] For $X<X'<1-(1-X)\cos^2\theta$, there are both transmitted and reflected modes. 
\item[B.] For $1-(1-X)\cos^2\theta<X'<1$, there is total internal reflection and no transmitted wave. As in a standard derivation of the Fresnel coefficients, one can use the result in case~A by formally allowing (in this case) $\cos\theta'>1$, with $\sin\theta'=(1-\cos^2\theta')^{1/2}$ imaginary.
\item[C.] For $X'>1$ the refractive index in the denser medium is imaginary, and there is no transmitted wave. In this case one has $\cos\theta'$ imaginary, with $\sin\theta'=(1-\cos^2\theta')^{1/2}>1$.
\end{description}
It is convenient to introduce parameters
\be
a=\sin\theta'/\sin\theta,
\qquad
c=\cos\theta/\cos\theta',
\label{ac}
\ee
which are real in case~A and one of which is imaginary in cases~B and~C, respectively. One may rationalize the expressions for the Fresnel coefficients such that $a,c$ appear only either squared or in the combination $ac$. Then in both cases~B and~C  $ac$ is imaginary, and these two cases can be treated together. Before rationalization, the Fresnel coefficients are
\be
{E''_R\over E_R}={c(1-a^2)\over(a+c)(1+ac)},
\qquad
{E''_L\over E_R}={-a(1-c^2)\over(a+c)(1+ac)}.
\label{Fresnel}
\ee

\subsection{Reflection and transmission}

The reflection coefficients may be calculated directly for case~A using (\ref{Fresnel}). One finds
\be
p_R={\cos2\theta+\cos2\theta'
\over
1+\cos2\theta\cos2\theta'},
\qquad
\cos2\theta'={X'-X+(1-X)\cos2\theta
\over1-X'}.
\label{pRA}
\ee
The total reflection coefficient is
\be
r_{\rm tot}={(1+\cos2\theta\,\cos2\theta')(\cos2\theta-\cos2\theta')^2
\over2\sin^2(\theta+\theta')(\sin2\theta+\sin2\theta')^2}.
\label{rc}
\ee
The angle $\theta'$ increases with increasing frequency, from $\theta'=0$ at $(\omega^2-\omega'^2_p)\sin^2\theta=\omega'^2_p-\omega^2_p$ to $\theta'\to\theta$ for $\omega\to\infty$. 

For $X'\ll1$ most of the energy is transmitted \citep{H85a}, so that this case is of little interest here. The degree of depolarization written down by \cite{WZM86} is reproduced by the high-frequency limit, $X,X'\to0$, of (\ref{pRA}), when one has $\theta'=\theta$ and $2\theta$ is identified as the angle of deflection. A different derivation of this limiting case is given in Appendix~B, and this alternative derivation shows that the result applies when the incident ray is not in the plane defined by the magnetic field and the normal to the boundary.

\subsection{Total internal reflection}

In cases~B and~C there is total internal reflection, so that one has $r_{\rm tot}=1$. After rationalization one finds
\be
p_R={(c^2+a^2)(1+c^2a^2)-4c^2a^2
\over
(c^2-a^2)(1-c^2a^2)},
\label{pRBC}
\ee
with either $a^2<0$ (case~B) or $c^2<0$ (case~C). In case~C, $c^2$ is negative, and it is convenient to write $c^2=-1/x$, with $x=(1-X)/(X'-1)=(\omega^2-\omega_p^2)(\omega'^2_p-\omega^2)$. One then has $a^2\sin^2\theta=1+x\cos^2\theta>0$. Thus in case~C, (\ref{pRBC}) simplifies to
\be
p_R={x-(1+3x)\sin^2\theta+2x\sin^4\theta
\over x+(1-x)\sin^2\theta}.
\label{pR2}
\ee
One finds that (\ref{pR2}) also applies in case~B, where $a^2$ and $x<-1/\cos^2\theta$ are negative. 

The two expressions (\ref{pRA}), for case~A, and (\ref{pR2}), for cases~B and~C, cover the entire range of angles and frequencies, $\omega>\omega_p$. The boundary between their regions of validity is $\theta'=0$ in (\ref{pRA}) and $x=-1/\cos^2\theta$ in (\ref{pR2}). Contour plots of $p_R$ as a function of $\omega/\omega_p$ and $\theta$ have similar forms for different $r$, provided the axes are scaled appropriately. An approximation to (\ref{pR2}) that exhibits this scaling is for $x\ll1$ and $\sin^2\theta\ll1$ when one has
\be
p_R\approx{1-3\xi(\omega\theta/\omega_p)^2
\over1+\xi(\omega\theta/\omega_p)^2},
\label{pR2a}
\ee
which applies for $1\ll\omega/\omega_p\ll\xi^{1/2}$, $\theta\ll1$.

\subsection{Depolarization coefficient}

\begin{figure}
\centerline{
\includegraphics[width=0.4\textwidth]{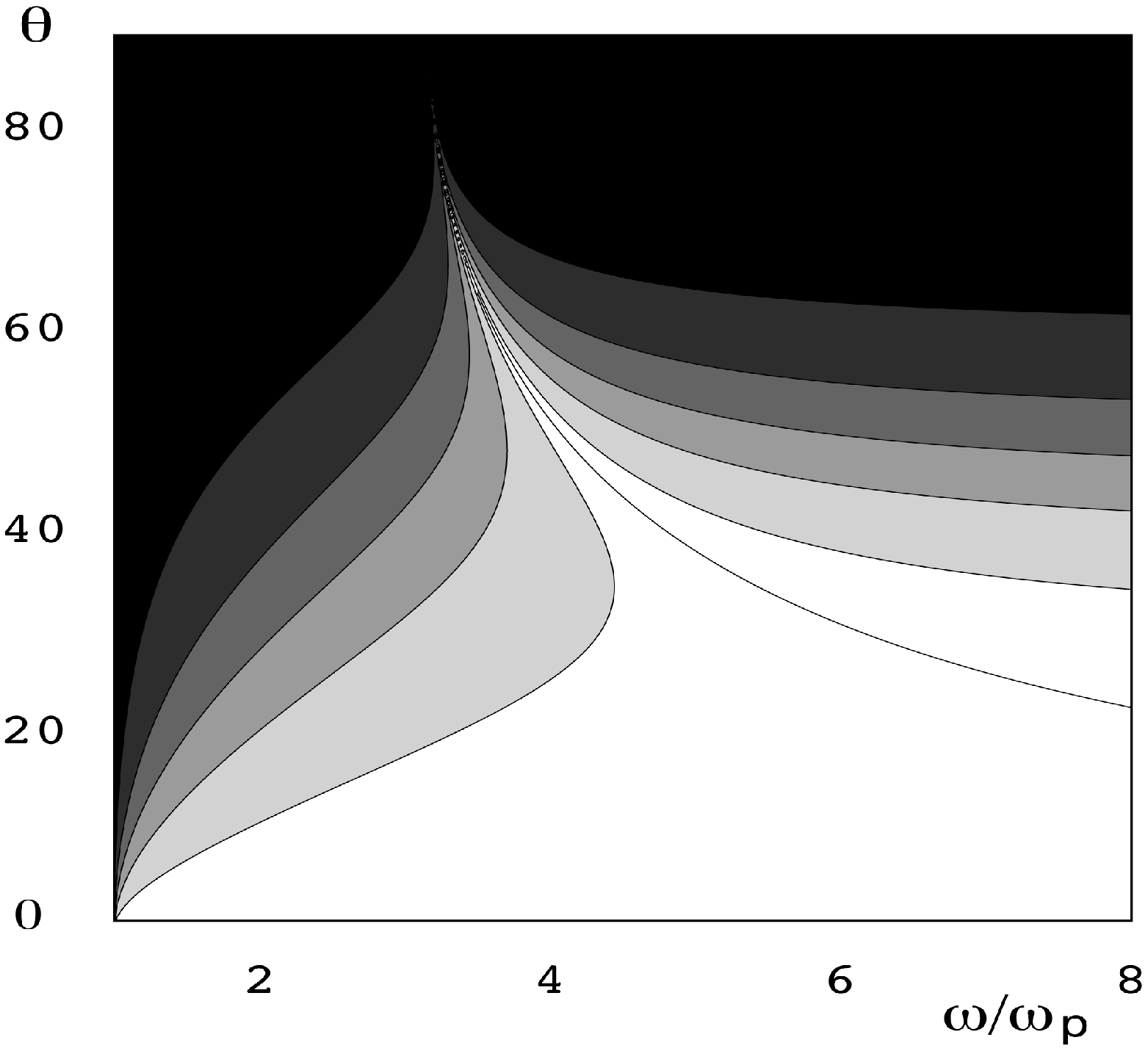}
\qquad
\includegraphics[width=0.4\textwidth]{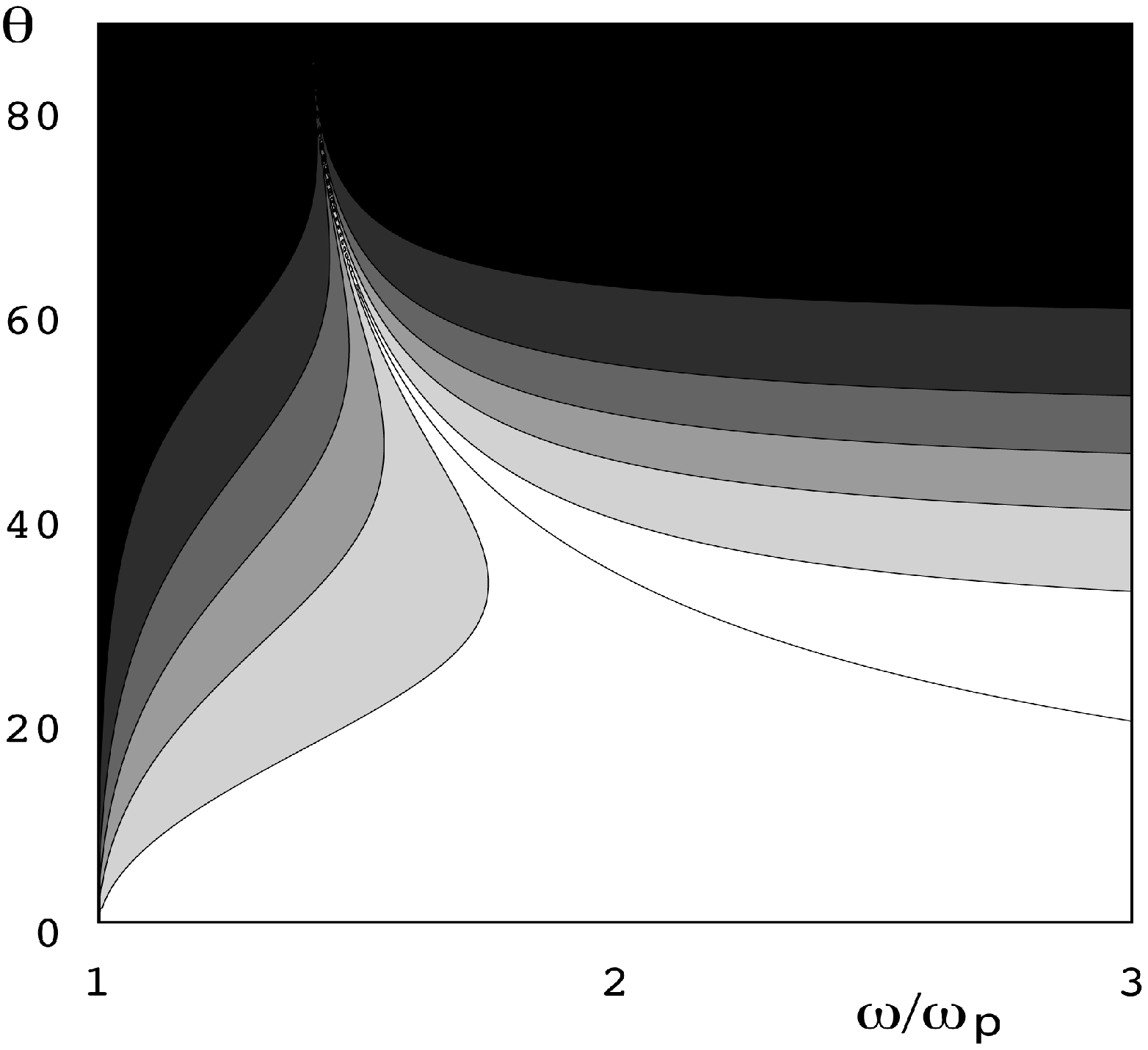}
}
\caption{The depolarization coefficient is shown as a contour plot on the $\omega/\omega_p$--$\theta$ plane  for $\xi=\omega'^2_p/\omega_p^2=10$ (left) and 2 (right). The contours are, from bottom to top, $p_R=0.8,0.4,0,-0.4,-0.8$. Thus the unshaded region correspond to depolarization coefficients $0.8<p_R<1$, and the dark region corresponds to $-1<p_R<-0.8$. The curve contained within the unshaded region, and between the contours where they converge on the upper right, corresponds to $p_R=1$; this curve separates total internal reflection, to its left, and partial reflection and partial transmission, to its right. The ranges of frequency for the two figures are chosen to emphasize the approximate scaling as a function of $\xi^{1/2}$.}
\label{fig:depol}
\end{figure}

Numerical results for the depolarization coefficient are shown as a contour plot in figure~\ref{fig:depol}. The depolarization coefficient is close to unity  for small angles of propagation (unshaded area), corresponding to large angles of incidence, and it is close to minus unity for large angles of propagation (dark area), corresponding to large angles of incidence. This implies that there is no significant reduction in polarization on reflection for reflection at sufficiently small angles, and that there is almost complete reversal of the polarization on reflection at sufficiently large angles. Strong depolarization occurs for reflection near the contour $p_R=0$. There are two sections of the contour $p_R=0$ in figure~\ref{fig:depol}, one in the region of total internal reflection and the other in the region where both reflection and transmission occur. Only the case of total internal reflection is likely to be important in practice, and then, for semiquantitative purposes, this contour may be approximated by (\ref{pR2a}), which corresponds roughly to a straight line from $\omega=\omega_p$, $\theta=0$ to $\omega=\omega'_p$, $\theta=45^\circ$. Any reflection that occurs near this line leads to almost complete depolarization. 

\begin{figure}
\centerline{
\includegraphics[width=0.4\textwidth]{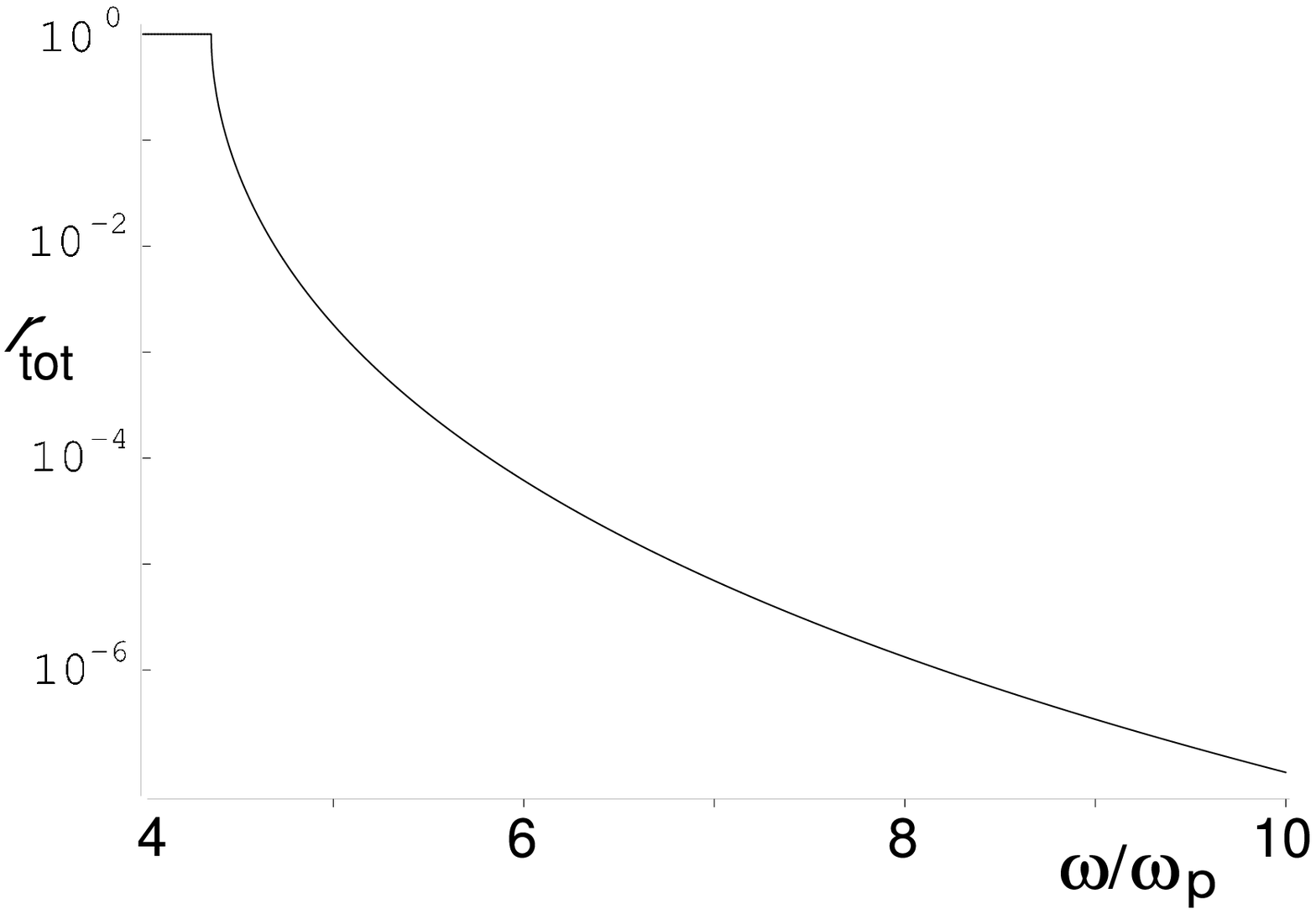}
\qquad
\includegraphics[width=0.4\textwidth]{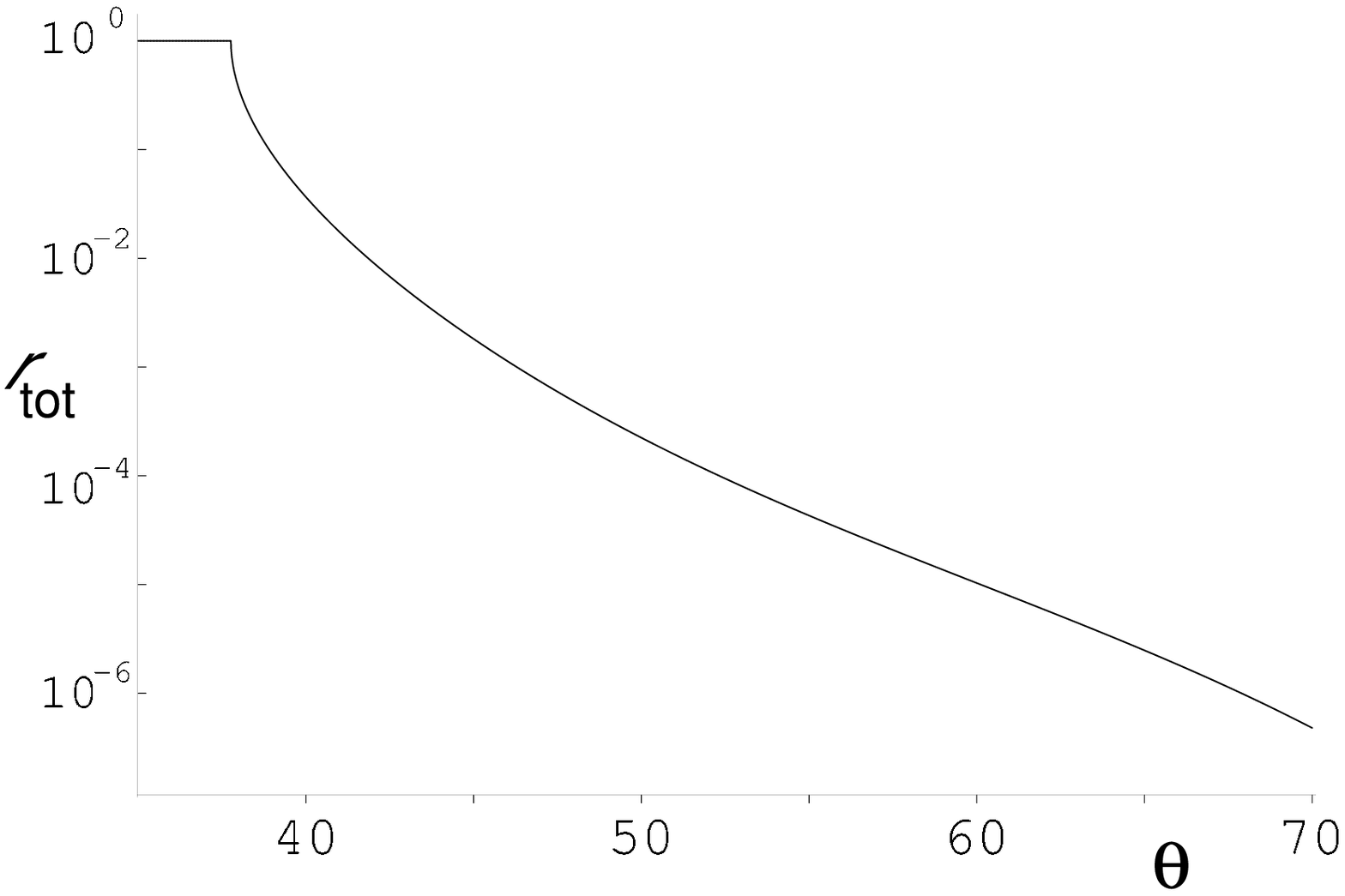}
}
\caption{The reflection coefficient $r_{\rm tot}$  is plotted for a density ratio $\xi=10$ is plotted as a function of $\omega/\omega_p$ for $\theta=45^\circ$ (left) and as a function of $\theta$ (in degrees) for $\omega/\omega_p=5$ (right).}
\label{fig:ref}
\end{figure}

The region where total internal reflection occurs is separated from the region where both reflection and transmission occur; in the two contour plots in figure~\ref{fig:depol} these regions are to the lower left and the upper right, respectively, of the curve corresponding to $p_R=1$. The region below this curve, where total internal reflection occurs, is restricted to smaller and smaller angles, $\theta$, as $\omega/\omega_p$ increases. Although such small-angle reflection might be effective in causing ducting, it is not effective in causing depolarization. Above the curve  $p_R=1$ in figure~\ref{fig:depol}, both reflection and transmission occur at the boundary. For reflection to be effective in causing depolarization in this case two conditions need to be satisfied: the depolarization coefficient should be close to zero, and the reflection coefficient, $r_{\rm tot}$, should not be too small.  The depolarization coefficient well to the right of the curve  $p_R=1$ in figure~\ref{fig:depol} is well approximated by setting $\theta'=\theta$ in (\ref{pRA}), which reproduces the formula given by \cite{WZM86}. In this regime, the contour line for $p_R=0$ asymptotes to $\theta\to45^\circ$ for $\omega/\omega_p\to\infty$, corresponding to a deflection through $2\theta=90^\circ$. As argued by \cite{WZM86}, it is difficult to account for the observed depolarization in terms of a single reflection with this depolarization coefficient, and multiple reflections are required to account for substantial depolarization. Each reflection decreases the reflected intensity by a factor equal to $r_{\rm tot}$, and if $r_{\rm tot}$ is very small multiple reflections lead to an extremely small intensity that can be of no interest in practice. The reflection coefficient is plotted in figure~\ref{fig:ref}. One has $r_{\rm tot}=1$ for total internal reflection, and figure~\ref{fig:ref} shows that $r_{\rm tot}$ falls off very rapidly in both frequency and angle above the threshold where transmission is allowed. Thus, in the regime well to the right of the curve  $p_R=1$ in figure~\ref{fig:depol} the intensity of the reflected component is orders of magnitude smaller than the transmitted intensity. It is concluded that effective depolarization requires total internal reflection, and that only the region to the left of the curve  $p_R=1$ in figure~\ref{fig:depol} is likely to be relevant in developing a model for depolarization.

\subsection{Validity of the model}

The model used here to calculate the depolarization coefficients corresponds to the limiting case of the magnetoionic theory in which the ratio of the electron cyclotron frequency to the plasma frequency is assumed negligibly small. The actual value of this ratio is thought to be small, of order $\Omega_e/\omega_p\approx0.1$ in type~III sources, but may be larger in type~I sources. Thus the formulae derived here are expected to be accurate only to of order 10\%, which should suffice for semiquantitative purposes. However, there are special cases where the approximation fails and might be misleading, specifically for $\omega-\omega_p\lessim\Omega_e$ and $\omega-\omega'_p\lessim\Omega_e$. In these cases, the refractive indices and the polarization vectors of the magnetoionic modes depend sensitively on the ratio of the frequency differences, $\omega-\omega_p$, $\omega-\omega'_p$, to $\Omega_e$. In particular, there is a range of frequencies above $\omega_p$ ($\omega'_p$) on the lower (higher) density side where the o~mode can propagate and the x~mode is evanescent. There is also a range around of frequencies $\approx\Omega_e$ around $\omega'_p$ in the denser medium where a transmitted component in the z~mode can propagate. 

The invalidity of the model for $\omega\approx\omega_p$ implies it cannot be used to treat any reflection that occurs near the point of emission. The x~mode exists only at frequencies above its cutoff, $\omega\gsim\omega_p+\Omega_e/2$, so that there can be no reflected $x$ mode for $\omega/\omega_p<\Omega_e/2\omega_p$. If one defines the reflection coefficients $r_{\rm ox},r_{\rm oo}$ for incident o~mode and reflected o~and x~modes, respectively, then one has $r_{\rm ox}=0$ below the x~mode cutoff. Moreover, there is a range of slightly higher frequencies where the refractive index of the x~mode is much smaller than that of the o~mode, and this mismatch implies  $r_{\rm ox}\ll r_{\rm oo}$. Thus the approximation in which the effect of the magnetic field on the refractive index is neglected is valid only for $\omega-\omega_p\gg\Omega_e$. This limitation is unimportant provided the first reflection occurs well above the source region, where both the o~and x~modes can propagate, and this is assumed to be the case in the following discussion. 

\begin{figure}
\centerline{
\includegraphics[width=0.9\textwidth]{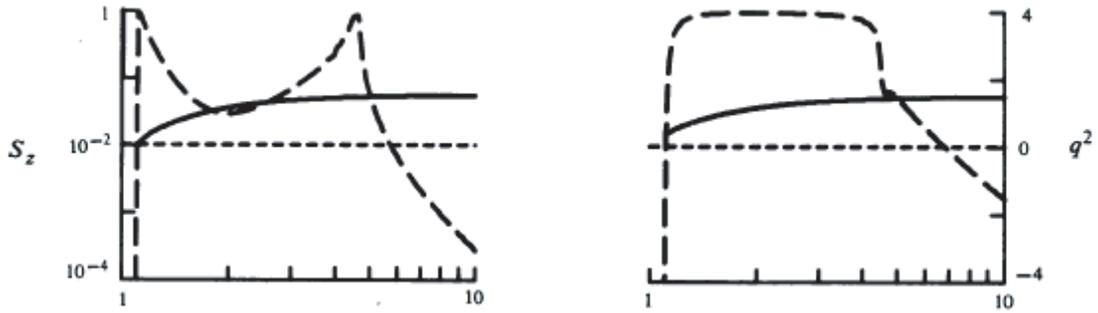}
}
\caption{The reflection coefficients ($S_z$ is the reflected Poynting flux along the $z$ axis for unity initial Poynting flux equal to unity) and the solution, $q^2$ of the Booker quartic equation are shown for an incident x~mode on a boundary with a density contrast $\omega'^2_p/\omega_p^2=8$ with $\Omega_e/\omega_p=0.2$ and $\theta=60^\circ$. The reflected x~and o~modes are shown on the left and right, respectively. [From Fig.~2 of Hayes (1985a).]}
\label{fig:hayes}
\end{figure}

The accuracy of the approximate model assumed here can be tested only by comparison with calculation using the exact magnetoionic expressions. Numerical calculations in the general case have been presented by  \cite{H85a,H85b}. An example of the exact case is illustrated in figure~\ref{fig:hayes}. This case is similar to that shown on the left in figure~\ref{fig:depol} for $\theta\approx60^\circ$. The exact result shown in figure~\ref{fig:hayes} is for an incident x~mode, and the results show that the polarization approximately reverses sense over most of the frequency range below the threshold for transmission: specifically, the reflected o~mode is much stronger than the reflected x~mode. However, this is not the case near $\omega=\omega_p$ and near the threshold for the appearance of a transmitted component. As already remarked, the approximate theory is invalid near $\omega=\omega_p$, and it fails to predict the correct behavior near $\omega=\omega_p$.  However, it does predict the sharp narrow peak with $p_R\approx1$ near the threshold for transmission. This suggests that the approximate model is valid for semiquantitative purposes except near $\omega=\omega_p$.

The calculations performed here are all for the case where the magnetic field, the incident ray and the normal to the boundary are coplanar. This limitation is unimportant and the formulae apply to the more general case, provided that the angle $\theta$ is reinterpreted appropriately. It is relatively simple to show this in the high-frequency limit (Appendix~B); for arbitrary angles $\theta,\phi$, the result implies by (\ref{pRA}), with $\theta'=\theta$, is reproduced with $2\theta$ interpreted as the angle of deflection of the ray in this more general case. The foregoing results are insensitive to the assumption of coplanarity, and their generalizations involve only a reinterpretation of the angle $\theta$.

\section{Depolarization of F~emission}

In applying the foregoing results to the interpretation of F~emission, it is important to distinguish between type~I emission and type~III emission, which have different characteristics. (The polarization of F type~II emission is not discussed separately here: it is similar to F type~III emission.) Type~I emission from sources near the central meridian experience little depolarization, and the degree of depolarization increases systematically with displacement of the source towards the limb. Type~III is never 100\% polarized, with a maximum polarization of about 70\%, with some bursts having a much lower polarization. A working hypothesis is that these differences are due in part to different density structures. Type~I emission is assumed to come from closed-loop regions where the magnetic field is relatively strong and the density structures are aligned along field lines whose direction changes only over distances of order the  major radius to the loop. Type~III emission is assumed to come from open field regions, where the magnetic field is weaker and the density structures are aligned along field lines that have a random component  about a mean orientation.

\subsection{Effect of refraction into the radial direction}

An important effect for any F~emission is an initial strong refraction towards the local direction of the outward density gradient, which for the purpose of discussion is assumed to be the radial direction. The theory of F~emission implies that the radiation is generated at a frequency that is above the local plasma frequency by fraction $f_0\approx3V_e^2/2v_\phi^2$, where $V_e$ is the thermal speed of electrons and $v_\phi$ is the phase speed of Langmuir waves, identified as the speed of the electrons exciting the emission. For flare-associated type~III emission this fraction is of order $0.01$, and is somewhat larger for storm associated type~III emission, and perhaps also for type~I emission. The refractive index at the point of emission is $n_0\approx\sqrt{2}f_0$. Let $\psi$ be the angle between the ray and the radial direction. Refraction implies that $n\sin\psi$ is constant along a ray, Hence, along a ray, $\psi$ varies so that it satisfies $(1-\omega_p^2/\omega^2)\sin^2\psi=2f_0^2$. For example, for $f_0=10^{-2}$ all the rays are confined to within $\psi\approx10^\circ$ of the radial direction when the plasma frequency has decreased by a few percent. For magnetic field lines roughly in the radial direction, it then follows that when a ray encounters a boundary the angle $\theta$ is determined primarily by the angle between the magnetic field and the radial direction.

Consider the implication in connection with the depolarization coefficients shown in figure~\ref{fig:depol}. F~emission is generated very near the left-hand boundary of the figure. Radial refraction causes it to move rapidly to smaller angles as the frequency increases. If the ray does not encounter a boundary, it moves further towards the lower right of the figure, and if there were no other effect, the ray would emerge within a few degrees of the radial direction. Either refraction or reflection off density irregularities must occur to produce the inferred relative broad angular patterns of escaping radiation. 

\subsection{Type~I emission}

The model of \cite{WZM86} accounts for most of the important features of the directivity and depolarization of type~I sources as they approach the limb. In the model, before encountering the boundary, the radiation is assumed to fill a cone of angles about the radial direction, and the magnetic field line that defines the boundary is at an angle $\theta$ to the radial. \cite{WZM86} argued that the depolarization must take place well above the source region, specifically where $\omega/\omega_p>2$, in order to be consistent with a constraint on the time delay between the o-mode and x-mode components. This model is adopted here, except that the reflection is attributed to the density jump at the surface of an overdense fiber, rather than to scattering by lower-hybrid waves. The important change is in the depolarization coefficient, which has different forms for these two types of reflection.

The depolarization coefficient for scattering by lower-hybrid waves corresponds to (\ref{pRA}) with $\theta'\to\theta$. An attractive feature of this model is that the depolarization depends only on the geometry, specifically, the angle through which the ray is deflected. It correctly predicts unpolarized radiation for $2\theta=90^\circ$, corresponding to a source on the limb. However, the model encounters a difficulty: averaging the reflected radiation over the assumed cone of angles makes it difficult to account for unpolarized type~I sources near the limb. This led \cite{WZM86} to assume multiple (at least three) reflections in order to account for the depolarization. For reflection at a sharp boundary, the same formula for the depolarization applies for $\omega\gg\omega'_p$. However, the reflection coefficient, $r_{\rm tot}$, is then very small, cf.\ figure~\ref{fig:ref}, and after three reflections the intensity would be negligibly small. To account for effective depolarization without a large reduction in intensity, the reflection must occur near or below the threshold for transmission to be allowed. The only plausible possibility is thus depolarization on total internal reflection.

Three constraints severely limit the possibilities: the need for total internal reflection, the inference from the time delay between the o~and x~mode component implying depolarization at $\omega/\omega_p\gsim2$, and the requirement that the deflection occur through roughly $90^\circ$ ($\theta\approx45^\circ$) for sources near the limb. These constraints can be satisfied simultaneously only for a relatively large density ratio, $\xi\approx10$, with the depolarization coefficient is then given by (\ref{pR2}). It is possible to have a relatively large depolarization for a single reflection. However, as in the model of \cite{WZM86}, for radiation in a cone of angles it is difficult to account for very small degrees of polarization with a single reflection. There is an additional effect that is absent in the model of \cite{WZM86}: the depolarization depends on frequency as well as angle. According to figure~\ref{fig:depol}, a given fiber reflecting radiation with a range of frequencies at a given angle has $p_R$ increasing with frequency; if $p_R=0$ occurs in this frequency range, then the lower frequency radiation should have the opposite handedness to that of the higher frequency radiation. Observation of such an effect would support the form of the model suggested here.

\subsection{Type~III emission}

In contrast to type~I burst, which are typically 100\% polarized, F emission in type~III  bursts is never 100\% polarized. In the framework of the present model, this requires that the radiation always experience at least one depolarizing reflection before escaping. 

The picture of the source region assumed here is a low-density background region, with overdense field-aligned structures acting as the walls of ducts \citep{D79}. For simplicity let us assume that all these structure have the same overdensity, $\xi$. The structures are also likely to have a distribution in angle, $\theta$, due to random wandering of the open field lines. The near-radial ducting suggests that the mean value of $\theta$ is relatively small: in the following discussion a mean $\theta\approx20^\circ$ is assumed. Type~III emission is assumed to be generated in the low-density region, to be refracted quickly towards the radial direction, to experience its first reflection off one of the overdense structures, to experience at least a few more reflections as it propagates outward, and to escape when it encounters no more structures capable of causing total internal reflection. The data suggest that both F and H emissions are ducted \citep{D79}, implying at escape occurs above the second harmonic plasma level, with $\omega/\omega_p>2$ at the escape point. The condition for total internal reflection to cease at the escape point is 
\be
(\omega/\omega_p)^2\sin^2\theta-\xi+\cos^2\theta=0.
\label{bound}
\ee
Assuming $\omega/\omega_p\approx2.5$ and $\theta\approx20^\circ$, (\ref{bound}) suggests that structures with $\xi\approx2$ are involved in ducting. 

Of particular importance for depolarization is the first reflection. According to figure~\ref{fig:depol} for $\xi=2$, for $\theta\approx20^\circ$ all reflections at $\omega/\omega_p\gsim1.5$ have $p_R>0.8$, implying that many such reflections would be required to reduce the polarization even to the highest observed values, $\approx70$\%. A possible model for depolarization of type~III bursts is that the first reflection occur at $1\ll\omega/\omega_p\lessim1.5$, and leads to a significant depolarization, and that two or more subsequent reflections occur, leading to further depolarization by lesser amounts. For the most highly polarized bursts, the model suggests $\omega/\omega_p\approx1.5$ with a further reflection at $\omega/\omega_p\approx2$ before escaping at $\omega/\omega_p\approx2.5$. Lower polarization is more plausibly attributed to the first reflection being closer to the source, leading to a greater depolarization, rather than to a greater number of subsequent reflections, which become increasing ineffective in causing depolarization.

It may be concluded that depolarization due to reflection off the walls of a duct can account for the observed polarization of F type~III bursts. The walls of an appropriate duct would consist of overdense structures with $\xi\approx2$ along magnetic field lines oriented at $\theta\approx20^\circ$ to the radial. The model suggests that all type~III bursts experience at least one reflection, with several further reflections likely before escape from the duct at $\omega/\omega_p>2$. Most of the depolarization occurs at the first reflection.

\subsection{Depolarization at a QT region}

Effective depolarization requires that radiation initially in one magnetoionic mode becomes a mixture of the two magnetoionic modes, and although reflection is one possibility, there is another possibility. Mode coupling is a weak effect except under two conditions: when the density gradients are very large, as assumed here, and near coupling points. In the absence of sharp density gradients, mode coupling in the corona is important only at so-called QT regions, which are coupling points where the angle, $\theta$, between the ray and the magnetic field passes through $90^\circ$. (QT is an abbreviation for `quasi-transverse' in now outdated jargon: in modern usage this corresponds to `quasi-perpendicular' when the polarization is better approximated by linear rather than circular.) When a QT region is encountered at sufficiently low frequencies, the radiation remains in the initial magnetoionic mode and the sense of polarization reverses, whereas for an encounter at sufficiently high frequencies, the sense of polarization is preserved,\citep{B61,Z70,M80}. Significant depolarization at a QT region occurs only at frequencies near the transition frequency between these limits \citep[e.g.][vol.\ II, p.\  300]{M80}. In principle, this provides an alternative depolarizing mechanism.

No model for depolarization of F~emission has been proposed based on QT regions. Any such model would encounter two major difficulties, based on the requirements that a QT region be encountered near the transition frequency. The geometry required for the rays to encounter an appropriate QT region does not fit with either the fiber model for type~I sources or the ducting model for type~III sources. Furthermore, there is no obvious reason why in any specific model a QT would be encountered near the transition frequency. 

\section{Conclusions}

In this paper it is assumed that depolarization of F~emission in type~I, II and~III bursts is due to reflection. In general, a wave in one magnetoionic mode incident on a sharp boundary leads to reflected waves in both modes, and to transmitted waves in both modes provided these waves can propagate in the denser (by a factor $\xi$) medium. The reflection model requires that the boundary have a thickness less than about the (free-space) wavelength of the radio waves.

The depolarization is treated in section~3 in terms of a depolarization coefficient, $p_R$, with $p_R=0$ corresponding to complete depolarization, and $p_R=\pm1$ corresponding to no change in polarization and reversal of (circular) polarization, respectively. A simplified form of magnetoionic theory is used to calculate $p_R$. Two different expressions are derived for $p_R$: (\ref{pRA}) applies when the transmitted waves can propagate, with the reflection coefficient then given by (\ref{rc}), and (\ref{pR2}) applies when total internal reflection occurs. Contour plots of $p_R$ as a function of $\omega/\omega_p$ and $\theta$ are shown in figures~\ref{fig:depol}: the general form is not sensitive to $\xi$, provided the frequency is scaled appropriately, as exhibited by the approximate form (\ref{pR2a}). The simplified form of magnetoionic theory reproduces relevant features found from exact calculations of reflection coefficients \citep{H85a,H85b}, except for $\omega\approx\omega_p,\omega'_p$, where the neglect of the magnetic field on the refractive indices of the magnetoionic waves is not justified. 

In applying the model to F~emission, it is assumed that the radial gradient in the corona density causes the ray to be strongly refracted towards the outward radial direction before it encounters a sharp boundary. The invalidity of the simplified model for $\omega\approx\omega_p$ is then unimportant, and the angle $\theta$ may be interpreted as approximately the angle between the magnetic field and the radial direction. It is found that when transmission is allowed the total reflection coefficient is very small except for a narrow range of parameters that is unlikely to be important in applications. On this basis, it is argued that effective depolarization requires total internal reflection.

The depolarization of type~I is attributed here to reflection off the overdense structures (fibers) invoked to account for the directivity \citep{BS77}. Type~I emission from sources near the central meridian must escape without encountering such a structure. For sources nearer the limb, the model proposed by \cite{WZM86} is adopted, except that the reflection (that these authors attributed to scattering off lower-hybrid waves) is assumed to be off a sharp boundary.  The time delay between the o~and x~mode components requires reflection well about the source region \citep{WZM86}, and for this to correspond to total internal reflection through a large angle (tending to $2\theta=90^\circ$ for a source at the limb) requires a large density contrast, $\xi\approx10$. Depolarization of F~emission in type~II and~III bursts can be explained in terms of reflection off ducts with a density contrast $\xi\approx2$. A ray then typically experiences two or more reflections before escaping, with the first reflection at $\omega/\omega_p\approx1.5$.

An important implication of the reflection model for depolarization is that there must be density structures with much more extreme properties than required by observations at non-radio wavelengths. There are few alternatives explanations for depolarization that would avoid this implication. One alternative is that proposed by \cite{WZM86} for type~I bursts: scattering off lower-hybrid waves generated through some appropriate instability. The lower-hybrid model requires that the lower-hybrid waves be present along a thin boundary layer that extends over a macroscopic distance and persists for many hours, and this is difficult to reconcile with the typical burstiness, in both space and time, of instabilities. Scattering by lower-hybrid waves has not been suggested for depolarization of F~emission in type~II and type~III bursts, and such a suggestion would encounter serious difficulties. In particular, the angles of deflection are thought to be relatively small during ducting, leading to a correspondingly small depolarization, whereas the depolarization is universal and relatively large for these bursts. Another possible depolarizing mechanism is associated with mode coupling at so-called QT regions. This mechanism has not been proposed for depolarization F~emission, and any model based on it would require very special conditions to apply. Another possibility that might be considered is that depolarization occurs very near to the source due to some unknown process that coverts o~mode waves into x~mode waves. Even if one postulated such a process, it would be inconsistent with the measured time delay between the o~and x~mode components in partially polarized type~I source, which imply depolarization at $\omega/\omega_p\gsim2$ \citep{WZM86}. Thus, the only seemingly viable depolarization mechanism for F~emission is reflection off density structures with large density contrasts and extremely sharp boundaries.

\acknowledgements
I thank Qinghuan Luo and Mike Wheatland for comments on the manuscript.

\appendix
\section{Condition for reflection}
\def\theequation{{\rm A}.\arabic{equation}}

Consider a ray encountering a boundary layer in which the plasma parameters vary as a function of only the coordinate $z$. Snell's law implies that $r=(k_x^2+k_y^2)^{1/2}c/\omega$ is a constant, and it is convenient to solve the magnetoionic dispersion equation (the `Booker quartic' equation) for the dependent variable to $q=k_zc/\omega$. A ray propagating from the lower density towards the higher density region encounters the reflection point for the o~mode before the reflection point for the x~mode. The condition for reflection to apply, in the sense that a continuous boundary layer is treated as a discontinuity, is that the skin depth for the o~mode in the region of evanescence between these reflection points exceed the distance between the reflection points. For $\Omega_e\ll\omega_p$, the frequency $\omega$ is approximately equal to the plasma frequency, $\omega_p$ near the reflection points, which occur at $1-X-Y-r^2\approx0$ and $1-X-r^2\approx0$, respectively. If $L$ is the characteristic length over which the density changes, the reflection points are separated by $\Delta z\approx(\Omega_e/\omega_p)L$. The skin depth is identified as the inverse of $(\omega/c){\rm Im}\,q$. The relevant approximate solution of the Booker quartic equation \citep[e.g.][]{B61} is for $\Omega_e/\omega\ll r$, and it may be approximated by
\be
{\rm Im}\,q=(\Omega_e/\omega_p)[(1-r^2)(1+r^2\cos^2\phi)]^{1/2},
\label{Imq}
\ee 
where $\psi$ is the angle between the magnetic field and the $z$-axis. Assuming the magnetic field lines are parallel to the surfaces of constant density implies $\cos\psi=0$, and then the condition for reflection becomes (\ref{reflection}).

\section{Depolarization at high frequencies}

The following derivation reproduces the reflection coefficient written down by \cite{WZM86}, and indicates how to relax the coplanarity assumption made in section~3.

Consider a model in which the magnetic field is along the 3~axis and the normal to the boundary is along the 1~direction. Let the incident wave be at polar angles $\theta,\phi$ relative to the magnetic field, with $\phi=0$ corresponding to the 1--3~plane. The incident wave is then along the direction $(\sin\theta\cos\phi,\sin\theta\sin\phi,\cos\theta)$. The reflected wave is along $(-\sin\theta\cos\phi,\sin\theta\sin\phi,\cos\theta)$. Right and left hand polarizations for the incident wave are
\be
{\bf e}_{R,L}={1\over\sqrt{2}}(\cos\theta\cos\phi\mp i\sin\phi,
\cos\theta\sin\phi\pm i\cos\phi,-\sin\theta).
\label{eRL}
\ee
Circular polarizations for the reflected waves, ${\bf e}''_{R,L}$ say, are given by $\cos\phi\to-\cos\phi$ in (\ref{eRL}). The reflection coefficients are in the ratio
\be
r_{RR}:r_{RL}=|{\bf e}^*_{R}\cdot{\bf e}'_{R}|^2:|{\bf e}^*_{R}\cdot{\bf e}''_{L}|^2,
\label{rs}
\ee
where an asterisk denotes complex conjugation. A straightforward calculation then gives
\be
p_R={(\cos^2\theta\cos^2\phi+\sin^2\phi)^2-(\sin^2\theta\cos^2\phi)^2
\over(\cos^2\theta\cos^2\phi+\sin^2\phi)^2+(\sin^2\theta\cos^2\phi)^2}.
\label{pR3}
\ee
The result (\ref{pR3}) simplifies in two special cases. For radiation incident in the 1--3~plane ($\phi=0$) one has $p_R=(\cos^4\theta-\sin^4\theta)/(\cos^4\theta+\sin^4\theta)$, and for radiation incident in the 1--2~plane ($\theta=\pi/2$) one has $p_R=(\sin^4\phi-\cos^4\phi)/(\sin^4\phi+\cos^4\phi)$. These two result are identical when written in terms of the angle of incidence (with the respect to the normal to the boundary), $\theta_{\rm inc}$, which is $\pi/2-\theta$ and $\phi$, respectively. Inspection of (\ref{pR3}) shows that the general result may also be expressed in this same form, that is
\be
p_R={\sin^4\theta_{\rm inc}-\cos^4\theta_{\rm inc}
\over\sin^4\theta_{\rm inc}+\cos^4\theta_{\rm inc}}
={2\cos\theta_{\rm def}
\over1+\cos^2\theta_{\rm def}},
\label{pR4}
\ee
where $\theta_{\rm def}=\pi-2\theta_{\rm inc}$ is the angle through which the ray is deflected. The result (\ref{pR4}) was written down using a more qualitative argument by \cite{WZM86}. 

The derivation in section~3 of the reflection coefficient off a sharp boundary is for $\phi=0$. The generalization to $\phi\ne0$ for simply by reinterpreting $2\theta$ as $\theta_{\rm def}$.

\newpage

\end{document}